\RequirePackage[2020-02-02]{latexrelease}
\documentclass[preprint,aps,tightenlines,showkeys,nofootinbib,superscriptaddress]{revtex4}
\usepackage{latexsym}
\usepackage[dvips]{color}
\usepackage[dvips]{graphicx}
\newcommand{\beq}{\begin{eqnarray}}
\newcommand{\eeq}{\end{eqnarray}}
\newcommand{\bea}{\begin{eqnarray*}}
\newcommand{\eea}{\end{eqnarray*}}
\newcommand{\eq}{eqnarray}

\newcommand{\ci}{\cite}

\newcommand{\De}{\Delta}

\newcommand{\la}{{\lambda}}
\newcommand{\La}{{\Lambda}}

\newcommand{\om}{{\omega}}

\newcommand{\pa}{{\partial}}
\newcommand{\no}{{\nonumber}}
\newcommand{\f}{\frac}
\newcommand{\ra}{\rightarrow}

\newcommand{\asy}{asymptotically}

\newcommand{\Ho}{Ho\v{r}ava}
\newcommand{\diff}{diffeomorphism}

\begin{document}

\preprint{arXiv:2309.13859v2 [hep-th]}

\title{
{Zero-Mass}
Rotating Spacetimes
in Four-Dimensional Ho\v{r}ava Gravity
}

\author{Mu-In Park\footnote{E-mail address: muinpark@gmail.com, Corresponding author}}
\affiliation{
Center for Quantum Spacetime, Sogang University,
Seoul, 121-742, Korea}

\author{Hyung Won Lee\footnote{E-mail address:hwlee@inje.ac.kr}}
\affiliation{
Institute of Basic Science and College of AI Convergence,
Inje University, 197 Inje-ro, Gimhae 50834, Korea}

\date{\today}

\begin{abstract}
We study a particular exact solution for rotating
spacetimes in four-dimensional
Ho\v{r}ava gravity, which has been proposed as a renormalizable
gravity model
without the ghost problem.
We show that the {zero-mass} Kerr spacetime
or the {zero-mass} Kerr-(A)dS
spacetime in Einstein gravity is an exact solution in
four-dimensional Ho\v{r}ava for an arbitrary IR Lorentz-violation parameter
$\la$, but with an appropriate cosmological constant. In particular, for the
{zero-mass} topological Kerr-AdS black hole solution with the hyperbolic
horizon topology or the {zero-mass} Kerr-dS cosmological solution with the spherical
horizon topology,
there exist the ergosphere and the
non-vanishing {(positive)}
Hawking temperature, which imply the existence of {\it negative} mass black holes as well as positive mass spacetimes, by losing its mass from the {zero-mass} ones via the Hawking radiation or Penrose process in the ergosphere.
\end{abstract}


\maketitle

\newpage

From the recent detections of gravitational waves for merging black holes or
neutron stars which are rotating generally, we opened the era of
strong gravity
(
UV) tests
of Einstein's general relativity (GR), beyond the usual weak gravity
(
IR) tests in the solar system \ci{Abbo}. In particular, we opened an era of testing the black hole physics, which has been just an academic subject. An exact solution of Einstein's equation for
a rotating black hole
was discovered by Kerr in 1963 \ci{Kerr} and has been an important
role in relativistic astrophysics as well as in theoretical physics.
In particular, due to the exactness and
uniqueness of the solution, even a slight departure from the solution
would indicate a modification of GR \ci{Abbo,Yune,LIGO:2021}, which could
be tested by analyzing the gravitational wave data in the near future.

It is known that GR is incomplete in UV where the quantum gravity effects are
important. Recently, a renormalizable 
gravity model without the ghost
problem, known as Ho\v{r}ava gravity, was proposed by abandoning Einstein's equal-footing treatment of space
and time through the anisotropic scaling dimensions $z>1$ \ci{Hora}.
A few years ago, an exact solution for the rotating black hole in
three-dimensional Ho\v{r}ava gravity has been found \ci{Park:1207},
but 
no exact solutions for rotating black holes in four-dimensional Ho\v{r}ava gravity has been found yet, except some approximate solutions for slowly rotating black holes \ci{Ghod,Lee:1008,Alie,Wang:1212}. In this paper, we consider the first step towards the four-dimensional rotating black hole solutions by finding a particular exact rotating solutions which could be a ``seed" solution for the more general solutions.

To this ends, we start by considering the ADM decomposition of the
metric
\begin{\eq}
\label{metric}
ds^2=-N^2 c^2 dt^2+g_{ij}\left(dx^i+N^i dt\right)\left(dx^j+N^j
dt\right)\
\end{\eq}
and the IR-modified Ho\v{r}ava action which reads
\begin{\eq}
S &= & \int dt d^3 x
\sqrt{g}N\left[\frac{2}{\kappa^2}\left(K_{ij}K^{ij}-\lambda
K^2\right)-\frac{\kappa^2}{2\nu^4}C_{ij}C^{ij}+\frac{\kappa^2
\mu}{2\nu^2}\epsilon^{ijk} R^{}_{i\ell} \nabla_{j}R^{\ell}{}_k
\right.
\nonumber \\
&&\left. -\frac{\kappa^2\mu^2}{8} R^{}_{ij}
R^{ij}+\frac{\kappa^2 \mu^2}{8(3\lambda-1)}
\left(\frac{4\lambda-1}{4}R^2-\Lambda_W R^{}+3
\Lambda_W^2\right)+\frac{\kappa^2 \mu^2 \om}{8(3\lambda-1)}
R^{}\right]\ , \label{horava}
\end{\eq}
where
\begin{\eq}
 K_{ij}=\frac{1}{2N}\left(\dot{g}_{ij}-\nabla_i
N_j-\nabla_jN_i\right)\
 \end{\eq}
is the extrinsic curvature,
\begin{\eq}
 C^{ij}=\epsilon^{ik\ell}\nabla_k
\left(R^{j}{}_\ell-\frac{1}{4}R^{} \delta^j_\ell\right)\
 \end{\eq}
is the Cotton tensor, and $\kappa,\lambda,\nu,\mu, \La_W$, $\om$
are constant parameters. The last term in (\ref{horava})
represents a ``soft" violation of the
``detailed balance" condition in \ci{Hora} and it modifies the IR
behaviors so that the Minkowskian flat
or Newtonian gravity limit may exist \ci{Keha,Park:0905}. Here, the Lorentz violations are introduced by the IR Lorentz-deformation parameter $\la$ $(\la \neq 1/3$; $\la=1$ is the GR case) and non-covariant deformations with higher-spatial derivatives (up to six orders, {\it i.e.}, $z=3$) \footnote{The gravity theory with a deformation parameter $\la_{DeWitt}$, which corresponds to $\la-1$ in this paper, has been first studied by DeWitt \ci{DeWi}. So, we may call the proposed theory as {\it Ho\v{r}ava-Lifshitz-DeWitt (HLD)} gravity, considering the earlier works of Lifshitz \ci{Lifs} and DeWitt \ci{DeWi}.} so that one can consider a renormalizable gravity theory without the ghost problem in UV but still recovers GR in IR with some appropriate limits \cite{Park:1508}.

The equations following from the variations of { a generic (non-projectable)} $N$ and $N^i$ are given by
\beq
&&\f{2}{\kappa^2}(K_{ij}K^{ij} -\lambda K^2) -\frac{\kappa^2\mu^2
\left[(\Lambda_W - \omega) R -3\Lambda_W^2 \right]}{8(1-3\lambda)}
-\frac{\kappa^2\mu^2 (1-4\lambda)}{32(1-3\lambda)}R^2+ \frac{\kappa^2}{2\nu^4} Z_{ij} Z^{ij}=0\, , \label{eom1} \\
&&\nabla_k(K^{k\ell}-\lambda\,Kg^{k\ell})=0\, ,\label{eom2}
\eeq
which correspond to the Hamiltonian and momentum constraints respectively, with
\beq
Z_{ij}\equiv C_{ij} - \frac{\mu \nu^2}{2} R_{ij}.
\eeq
 The equations of motion from variation of $\delta g^{ij}$ are quite messy and given by \ci{Lu,Kiri,Alie,Lee:1110,Deve:2018}
\beq
\frac{2}{\kappa^2}E_{ij}^{(1)}-\frac{2\lambda}{\kappa^2}E_{ij}^{(2)}
+\frac{\kappa^2\mu^2(\Lambda_W-\omega)}{8(1-3\lambda)}E_{ij}^{(3)}
+\frac{\kappa^2\mu^2(1-4\lambda)}{32(1-3\lambda)}E_{ij}^{(4)}
-\frac{\mu\kappa^2}{4\nu^2}E_{ij}^{(5)}
-\frac{\kappa^2}{2\nu^4}E_{ij}^{(6)}=0, \label{eom3}
\eeq
where
\bea
E_{ij}^{(1)}&=& N_i \nabla_k K^k{}_j + N_j\nabla_k K^k{}_i -K^k{}_i
\nabla_j N_k-
   K^k{}_j\nabla_i N_k - N^k\nabla_k K_{ij}\no\\
&& - 2N K_{ik} K_j{}^k
  -\frac{1}{2} N K^{k\ell} K_{k\ell}\, g_{ij} + N K K_{ij} + \dot K_{ij}
\,,\no \\
E_{ij}^{(2)}&=& \frac{1}{2} NK^2 g_{ij}+ N_i \pa_j K+
N_j \pa_i K- N^k (\pa_k K)g_{ij}+  \dot K\, g_{ij}\,,\no\\
E_{ij}^{(3)}&=&N\Big(R_{ij}- \frac{1}{2}R g_{ij}+\frac{3}{2}
\frac{\Lambda_W^2}{\Lambda_W-\omega} g_{ij}\Big)-(
\nabla_i\nabla_j-g_{ij}\nabla_k\nabla^k)N\,,\no\\
E_{ij}^{(4)}&=&NR\Big(2 R_{ij}-\frac{1}{2}R g_{ij}\Big)- 2
\big(\nabla_i\nabla_j-g_{ij}\nabla_k\nabla^k\big)(NR)\,,\no\\
E_{ij}^{(5)}&=&\nabla_k\big[\nabla_j(N Z^k_{~~i}) +\nabla_i(N
Z^k_{~~j})\big]  -\nabla_k\nabla^k(N Z_{ij})
-\nabla_k\nabla_\ell(N Z^{k\ell})g_{ij}\,, \no \\
E_{ij}^{(6)}&=&-\frac{1}{2}N Z_{k\ell}Z^{k\ell}g_{ij}+
2NZ_{ik}Z_j^{~k}-N(Z_{ik}C_j^{~k}+Z_{jk}C_i^{~k})
+NZ_{k\ell}C^{k\ell}g_{ij}\no\\
&&-\frac{1}{2}\nabla_k\big[N\epsilon^{mk\ell}
(Z_{mi}R_{j\ell}+Z_{mj}R_{i\ell})\big]
+\frac{1}{2}R^n{}_\ell\, \nabla_n\big[N\epsilon^{mk\ell}(Z_{mi}g_{kj}
+Z_{mj}g_{ki})\big]\no\\
&&-\frac{1}{2}\nabla_n\big[NZ_m^{~n}\epsilon^{mk\ell}
(g_{ki}R_{j\ell}+g_{kj}R_{i\ell})\big]
-\frac{1}{2}\nabla_n\nabla^n\nabla_k\big[N\epsilon^{mk\ell}
(Z_{mi}g_{j\ell}+Z_{mj}g_{i\ell})\big]\no\\
&&+\frac{1}{2}\nabla_n\big[\nabla_i\nabla_k(NZ_m^{~n}\epsilon^{mk\ell})
g_{j\ell}+\nabla_j\nabla_k(NZ_m^{~n}\epsilon^{mk\ell})
g_{i\ell}\big]\no\\
&&+\frac{1}{2}\nabla_\ell\big[\nabla_i\nabla_k(NZ_{mj}
\epsilon^{mk\ell})+\nabla_j\nabla_k(NZ_{mi}
\epsilon^{mk\ell})\big]
-\nabla_n\nabla_\ell\nabla_k
(NZ_m^{~n}\epsilon^{mk\ell})g_{ij}\,. \\
\eea

For non-rotating spherically symmetric solutions, one can get the exact
solutions 
by considering the appropriate metric ansatz
which depends only on the radial coordinate $r$ and the equations of motion for the resulting reduced
action \ci{Keha,Park:0905,Lu,Kiri}. However, for rotating axisymmetric
solutions, the metric functions depend on the polar angle $\theta$ as well as
the radial coordinate $r$ and the resulting equations of motion are non-linear partial differential
equations (PDE), contrast to the non-linear ordinary differential equations
(ODE) for the non-rotating cases, so that finding the exact solutions for the
rotating cases are formidable tasks and has not been succeeded yet \footnote{In three dimensions, due to the absence of the polar angle even for rotating black holes, one can get the (coupled, non-linear) ODE for the reduced equations of motion and the exact solution can be obtained \ci{Park:1207}.}, except
the approximate solutions for slowly rotating black holes \ci{Lee:1008,Alie,Wang:1212}.

In this paper, rather than obtaining the exact solutions for rotating black holes from the reduced action, we consider trial solutions so that the full equations of motion (\ref{eom1}-\ref{eom3}) are satisfied with an {\it arbitrary rotation} parameter $a$, by some inspections. For this purpose, we first note that
the well-known Kerr black hole solution in asymptotically flat spacetime  \ci{Kerr} can not be a solution of Ho\v{r}ava gravity  trivially \ci{Lee:1110}, which implies that some important deformations in the Kerr solution are needed beyond the GR limit. But, remarkably, we find that the {\it {zero-mass}} Kerr metric in the Boyer-Lindquist coordinates \cite{Boyer}
\beq
ds^2_{Kerr}=-dt^2 +\f{r^2 +a^2 \mbox{cos}^2\theta}{r^2+a^2} dr^2+(r^2 +a^2 \mbox{cos}^2\theta)d \theta^2 + (r^2+a^2) \mbox{sin}^2\theta d \phi^2
\label{masslessKerr}
\eeq
\textrm{}{\it could} be a solution of Ho\v{r}ava gravity since, for the
Kerr metric, all the curvature quantities
$R_{ij}, K_{ij}, C_{ij}, {\cdots, \it etc.}$ are proportional to the
black hole mass $M$ and hence they all vanish for the {zero-mass} black hole case. Actually, one can easily check that all the
equations of (\ref{eom1}-\ref{eom3}) are satisfied for $\La_W=0$,
{\it i.e.,} the asymptotically flat space-time, from the relation of
the cosmological constant in GR, $\La_{GR}=-3 \La_W^2 c^2/2 \om=0$ \cite{Park:1508},
but
for ``arbitrary" $\la$ and all other parameters in four-dimensional Ho\v{r}ava gravity (\ref{horava}) \footnote{In this paper, we do not consider the terms which depend on $a_i \equiv \partial_i N /N$ since{, in the non-projetable case $N=N(t, {\bf x})$,} those will change the dynamical degrees of freedom and also the IR (which reduces to a {\it scalar-tensor} theory, not GR) as well as UV behaviors a lot from those of the standard action (\ref{horava}) \cite{Blas:2009}. However, it is interesting to note that, for the low energy limit and the asymptotically {\it flat} space-time, the same {zero-mass} Kerr solution (\ref{masslessKerr}) satisfies the extended equations of motions with $a_i$ terms also
{since $a_i$ terms vanish trivially due to its projectable form $N=N(t)$
 (see also footnote 5)}.
 }. So, we find an asymptotically flat vacuum ({zero-mass}) rotating solution for Ho\v{r}ava gravity. But the solution has no black hole horizons and so it is not the desired solution for our purpose of finding rotating black hole solutions with horizons. \footnote{In \cite{Gibb:2017}, the {zero-mass} Kerr metric (\ref{masslessKerr}) is interpreted as a {\it wormhole} solution sourced by a negative tension ring.}

However, we note that there is one instance that the {zero-mass} spacetime solution can have black hole horizons, {\it i.e.}, the AdS black holes with the hyperbolic horizon topology $(k=-1)$ \ci{Mann:9607}, which are called {\it topological} AdS black holes. So, let us consider the {\it {zero-mass}} $Kerr-AdS_4$ ($KAdS_4$) metric with the hyperbolic horizon topology \ci{Klem} as a ``trial" solution
\beq
\label{masslessKerrhyper}
ds^2_{KAdS_4}=-\frac{\rho^2 \Delta_\theta \Delta_r}{\Xi^2 \Sigma^2} dt^2+\frac{\rho^2}{\Delta_r}dr^2+ \frac{\rho^2}{\Delta_\theta} d \theta^2
+\frac{\Sigma^2 \mbox{sinh}^2\theta}{\Xi^2 \rho^2} \left(d \phi-\frac{\La a}{3 \Xi} dt\right)^2,
\eeq
or the ADM metric (\ref{metric}) with
\begin{equation}
N^2 = \frac{\rho^2 \Delta_\theta \Delta_r}{\Xi^2 \Sigma^2}, \,\,
g_{rr} = \frac{\rho^2}{\Delta_r}, \,\,
g_{\theta\theta} = \frac{\rho^2}{\Delta_\theta}, \,\,
g_{\phi\phi} = \frac{\Sigma^2 \mbox{sinh}^2\theta}{\Xi^2 \rho^2 }, \,\,
N^{\phi} = - \frac{\La a}{3 \Xi}\,\,.
\label{KAdS4}
\end{equation}
Here,
\begin{eqnarray}
\rho^2 &=& r^2 + a^2 \mbox{cosh}^2 \theta, \nonumber \\
\Delta_\theta &=& 1-\f{\La a^2 \mbox{cosh}^2 \theta}{ 3},\nonumber \\
\Delta_r &=& \left( r^2 + a^2 \right) \left(-1 -\f{\La r^2}{3}\right), \no \\
\Xi &=&1-\f{\La a^2 }{ 3},  \nonumber \\
\Sigma^2 &=& \left( r^2 + a^2 \right) \rho^2 \Xi,
\end{eqnarray}
and $\La$ is a {\it negative} undetermined (cosmological) constant parameter ($\La<0$).

Another interesting instance with horizons is the ${zero-mass}~ Kerr-dS_4$ ($KdS_4$)
metric with the spherical horizon topology $(k=+1)$ \ci{Cart:73,Gibb:1977,Park:2001}
\footnote{It is interesting to note that the {zero-mass} Kerr metric (\ref{masslessKerr}) is
{in} the {\it projectable} form $N=N(t)$, though
not possible for
the {zero-mass} Kerr-(A)dS metric {in} (\ref{masslessKerrhyper})
{or} (\ref{masslessKerr_dS}), {which is in the genuine non-projectable form $N=N(t, {\bf x})$}.}
\beq
\label{masslessKerr_dS}
ds^2_{KdS_4}=-\frac{\bar{ \rho}^2 \bar{\Delta}_\theta \bar{\Delta}_r}{\bar{\Xi}^2 \bar{\Sigma}^2} dt^2+\frac{\bar{\rho}^2}{\bar{\Delta}_r}dr^2+ \frac{\bar{\rho}^2}{\bar{\Delta}_\theta} d \theta^2
+\frac{\bar{\Sigma}^2 \mbox{sin}^2\theta}{\bar{\Xi}^2 \bar{\rho}^2}
\left(d \phi-\frac{\La a}{3 \bar{\Xi}} dt \right)^2,
\eeq
or the ADM metric (\ref{metric}) with
\begin{equation}
N^2 = \frac{\bar{ \rho}^2 \bar{\Delta}_\theta \bar{\Delta}_r}{\bar{\Xi}^2 \bar{\Sigma}^2} , \,\,
g_{rr} = \frac{\bar{\rho}^2}{\bar{\Delta}_r}, \,\,
g_{\theta\theta} = \frac{\bar{\rho}^2}{\bar{\Delta}_\theta}, \,\,
g_{\phi\phi} = \frac{\bar{\Sigma}^2 \mbox{sin}^2\theta}{\bar{\Xi}^2\bar{\rho}^2} , \,\,
N^{\phi} = - \frac{\La a}{3 \bar{\Xi}}\,\,.
\label{KdS4}
\end{equation}
Here,
\begin{eqnarray}
\bar{\rho}^2 &=& r^2 + a^2 \mbox{cos}^2 \theta, \nonumber \\
\bar{\Delta}_\theta &=& 1+\f{\La a^2 \mbox{cos}^2 \theta}{ 3},\nonumber \\
\bar{\Delta}_r &=& \left( r^2 + a^2 \right) \left(1 -\f{\La r^2}{3}\right), \no \\
\bar{\Xi}&=&1+\f{\La a^2 }{ 3},  \nonumber \\
\bar{\Sigma}^2 &=& \left( r^2 + a^2 \right) \bar{\rho}^2 \bar{\Xi},
\end{eqnarray}
and $\La>0$.
Note that the $KdS_4$ metric (\ref{masslessKerr_dS}) can be obtained from the $KAdS_4$ metric (\ref{masslessKerrhyper}) by the analytic continuation \ci{Klem}
\beq
t \ra -it,~ r \ra -ir,~ \theta \ra -i\theta,~ \phi \ra \phi,~a \ra -i a,~ \La \ra -\La.
\label{analytic}
\eeq

It is interesting to note that both (\ref{masslessKerrhyper}) and
(\ref{masslessKerr_dS}) have the ``ergospheres" {\it outside} the rotating black hole's outer event horizon or {\it inside} the rotating cosmological horizon, where
$g_{tt}>0$, {\it i.e.},
\beq
{g_{tt}}_{ (KAdS_4)}&=&\left(1+\f{\La r^2}{3}-\f{\La}{3}a^2 \rm{sinh}^2  \theta \right)\left(1-\f{\La a^2}{3} \right)^{-2},\\
{g_{tt}}_{ (KdS_4)}&=&\left(-1+\f{\La r^2}{3}+\f{\La}{3}a^2 \rm{sin}^2  \theta \right)\left(1+\f{\La a^2}{3} \right)^{-2}
\eeq
are positive. Thus, in the region (see Figs. 1--3)
\beq
&&r_H < r<\sqrt{-\f{3}{\La}+a^2 \rm{sinh}^2 \theta} \equiv r_{\rm erg (H)},\label{ergo_KAdS}\\
&&r_C < r<\sqrt{\f{3}{\La}-a^2 \rm{sin}^2 \theta}\equiv r_{\rm erg (C)} \label{ergo_KdS}
\eeq
the positive energy extraction from rotating black holes or cosmological horizons are possible via Penrose process \ci{Penr}.

\begin{figure}
\includegraphics[width=7cm,keepaspectratio]{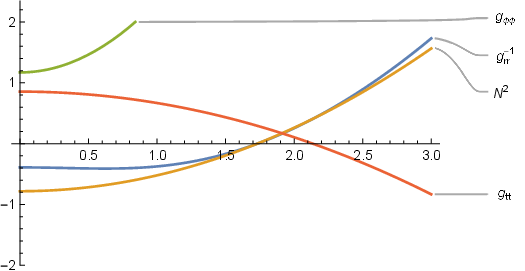}
\qquad
\includegraphics[width=7cm,keepaspectratio]{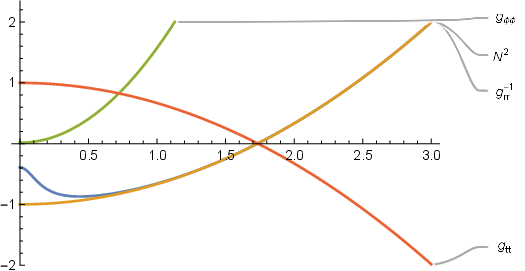}
\caption{Plots of $g_{rr}^{-1}(r)$ (blue), $N^2 (r)$ (orange), $g_{\phi \phi}$ (green), and $g_{tt}$ (red) for the {zero-mass} $KAdS_4$ metric with the hyperbolic horizon topology (\ref{KAdS4}). Here, we plotted $\Lambda=-1, \theta=\pi/3$ with $a=1$ (left) and $a=0.1$ (right) for a comparison. The left panel shows the existence of an ergosphere (\ref{ergo_KAdS}) {\it outside} the event horizon $r_H=\sqrt{3/(-\La)}$. The right panel shows the shrunk ergosphere as the rotation parameter $a$ approaches to zero. }\label{fig:metric}
\end{figure}

\begin{figure}
\includegraphics[width=7cm,keepaspectratio]{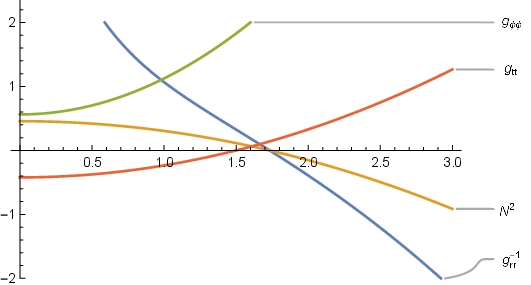}
\qquad
\includegraphics[width=7cm,keepaspectratio]{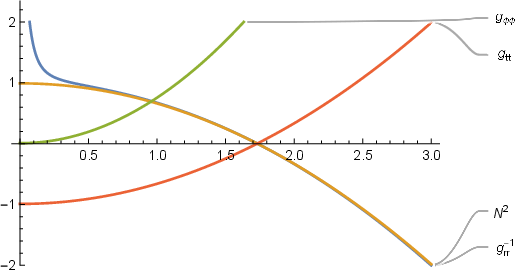}
\caption{Plots of $g_{rr}^{-1}(r)$ (blue), $N^2 (r)$ (orange), $g_{\phi \phi}$ (green), and $g_{tt}$ (red) for the {zero-mass} $KdS_4$ metric with the spherical horizon topology (\ref{KdS4}). Here, we plotted $\Lambda=1, \theta=\pi/3$ with $a=1$ (left) and $a=0.1$ (right). The left panel shows the existence of an ergosphere (\ref{ergo_KdS}) {\it inside} the cosmological horizon $r_C=\sqrt{3/\La}$. The right panel shows the shrunk ergosphere as the rotation parameter $a$ approaches to zero. }\label{fig:metric}
\end{figure}

\begin{figure}
\includegraphics[width=8cm,keepaspectratio]{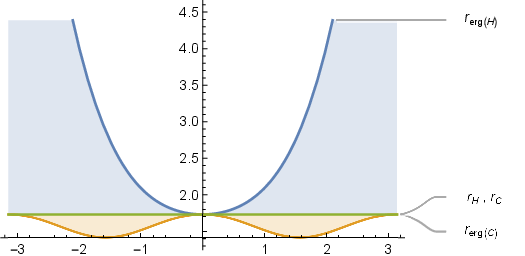}
\caption{Plots of the ergospheres for {zero-mass} $KAdS_4$ metric (\ref{KAdS4}) with the hyperbolic horizon topology
(blue region) and {zero-mass} $KdS_4$ metric (\ref{KdS4}) with the spherical
horizon topology (orange region). Here, we plotted $\La=\pm 1, a=1$ and
the green line denotes the black hole horizon $r_H=\sqrt{3/(-\La)}$ or
the cosmological horizon $r_C=\sqrt{3/\La}$, and the blue and
orange lines denote the corresponding ergosurfaces $r_{\rm erg (H)}$
and $r_{\rm erg (C)}$.
}\label{fig:metric}
\end{figure}

Now, we note that, for the solution (\ref{masslessKerrhyper}) or (\ref{masslessKerr_dS}), in contrast to the asymptotically flat rotating solution (\ref{masslessKerr}), it is not a trivial matter to compute the full equations of motion  (\ref{eom1}-\ref{eom3}). But with the help of computational programs, like {\it Maple} or {\it Mathematica} \footnote{In {\it Mathematica}, in order to have a fast computation for the laptop, we replace ${\rm cos}^{2} \theta=u(\theta)$ and ${\rm sin}^2 \theta =1-u(\theta)$ and also input the Ricci tensor and curvature scalar in their simplest forms (again with function $u(\theta)$). (We would like to thank D.~O.~Devecioglu for pointing out this.) In {\it Maple}, on the other hand, there are no troubles with cosines, sines.}, we have checked that the full equations (\ref{eom1}-\ref{eom3}) are satisfied if $\La$ satisfies the algebraic equation
\beq
\La^2+6 \La (\om-\La_W) +9 \La_W^2=0,
\eeq
which gives two solutions
\beq
\La=-3 \left[(\om-\La_W) \pm  \sqrt{\om (\om-2 \La_W)}\right]. \label{Lambda_eq}
\eeq
Here, $`-$' and $`+$' signs correspond to the GR and non-GR branches respectively since they behave as
\begin{\eq}
\La=
\left\{
\begin{array}{lll}
\La_{GR}+{\cal O}(\La_W^3/\om^2) \\
6 (\La_W-\om)-\La_{GR}+{\cal O}(\La_W^3/\om^2)
\end{array}
\right.
\end{\eq}
for the GR limit $|\om| \gg |\La_W|$, with the cosmological constant in
GR, $\La_{GR}=-3 \La_W^2 c^2/2 \om$ \ci{Park:1508}. Note that the \Ho~gravity action (\ref{horava}) is well defined only for the asymptotically AdS case ($\La_{GR}<0$) \cite{Hora,Lu} and the action for the asymptotically dS case ($\La_{GR}>0$) is given by the analytic continuations, $\mu \ra -i \mu, \om \ra -\om$ \cite{Park:0905}.


In order to understand this rather simple result from the messy equations (\ref{eom1}-\ref{eom3}), we have looked at some more details of the computations and found the following elegant behaviors of the solution
\beq
R_{ij}=\f{2}{3} \La g_{ij}, ~R=2 \La,~K_{ij}=0
\label{3D_eq}
\eeq
so that $\nabla_k R^{ij}=\nabla R=C^{ij}=0$ and the messy equations (\ref{eom1}-\ref{eom3}) become much simpler. In particular, the Hamiltonian constraint equation (\ref{eom1}) reduces to
\beq
\f{\kappa^2 \mu^2}{24 (1-3 \la)} [\La^2+6 \La (\om-\La_W) +9 \La_W^2 ]=0,
\eeq
which is consistent with (\ref{Lambda_eq}), while the momentum constraint
equation (\ref{eom2}) is trivially satisfied. We note that the solution
 (\ref{Lambda_eq}) does not depend on the IR Lorentz-deformation parameter
 $\la$, unless we consider the singular case of $\la=1/3$ which needs a
 separate consideration \cite{Park:2011,Kiri,Deve:2020}. The above result
 (\ref{3D_eq}) indicates that the spatial section of the {zero-mass} $Kerr-(A)dS_4$
  metric (\ref{masslessKerrhyper}) is {\it maximally symmetric}
 as in the three-dimensional Einstein gravity with a cosmological constant
 $\La/3$. Moreover, it is an interesting fact that the extrinsic curvature,
 which is symmetrized in the indices, vanishes
 $K_{ij}=0$ (\ref{3D_eq}) though its non-symmetrized components do not vanish, $\nabla_i N_j \neq0$.

It is important to note that, in contrast to the {zero-mass} Kerr solution with the vanishing cosmological constant (\ref{masslessKerr}), there is a black hole horizon $r_H$ for the \asy~AdS ($\La<0$) solution with the hyperbolic horizon topology (\ref{masslessKerrhyper}) or a cosmological horizon $r_C$ for the \asy~dS ($\La>0$) solution with the spherical horizon topology (\ref{masslessKerr_dS}) as
\beq
r_{H}=\sqrt{\f{3}{-\La}},~r_{C}=\sqrt{\f{3}{\La}},
\label{horizon}
\eeq
where $\De_r=0$. And also there are the non-vanishing Hawking temperature
\beq
T_H&=&\f{\De_r'|_{r_H}}{4 \pi (r_H^2 +a^2)}=\f{\sqrt{-\La}}{2 \pi \sqrt{3}},
\label{Hawking_Temp_H}\\
T_C&=&\f{|\De_r'|_{r_C}}{4 \pi (r_C^2 +a^2)}=\f{\sqrt{\La}}{2 \pi \sqrt{3}},
\label{Hawking_Temp_C}
\eeq
which are {\it positive} for $KdS_4$ as well as $KAdS_4$ cases, following the usual convention of \cite{Gibb:1977}.

Note that there is no dependence on the
{ rotation} parameter $a$ in (\ref{horizon})
and the final result of (\ref{Hawking_Temp_H}) or (\ref{Hawking_Temp_C}).
In GR, this fact may indicate an intimate connection of the rotating
metric (\ref{KdS4}) with the non-rotating
metric through some complicated coordinate transformations \cite{Cart:73,Henn} which
involve the time coordinate as well as the space coordinates.
In \Ho~gravity,
it might be thought to be ``improbable" due to lack of the {\it full} \diff. However, as we can see below, this is not correct and actually the transformation is still possible within the {\it foliation-preserving} \diff~(${\it Diff}_{\cal F}$) in \Ho~gravity \cite{Hora}:
\begin{\eq}
\delta_{\xi} t&=&-{\xi}^{0}(t),~~ \delta_{\xi} x^{i}=-\xi^{i}(t,{\bf x}),\label{deltx_Horava}\\
\delta_{\xi} N&=&(N{\xi}^{0})_{,0}+\xi^{k}\nabla_{k}N\label{delN3},\\
\delta_{\xi}{N_{i}}&=&{\xi}^{0}{}_{,0}N_{i}+\xi^{j}{}_{,0}g_{ij}+\nabla_{i}\xi^{j}N_{j}
+N_{i,0}{\xi}^{0}+\nabla_{j}N_{i}\,\xi^{j}\label{delNi3},\\
\delta_{\xi}{g_{ij}}&=&\nabla_{i}\xi^{k}g_{kj}+\nabla_{j}\xi^{k}g_{ki}
+g_{ij,0}{\xi}^{0}.\label{delg3}
\end{\eq}

To see this, we start from the {\it {zero-mass}} non-rotating spherically-symmetric black hole solution with the hyperbolic horizon topology in \Ho~gravity for the case of ``$\la=1$" \ci{Keha,Park:0905,Park:1508,Lu}
\begin{\eq}
  d\hat{s}^2=-\hat{N}(\hat{r})^2 dt^2+\frac{d\hat{r}^2}{\hat{f}(\hat{r})}+\hat{r}^2
\left(d\hat{\theta}^2+\sinh^2\hat{\theta} d\hat{\phi}^2\right)
\label{Static_metric}
\end{\eq}
with
\begin{\eq}
\hat{N}^2=\hat{f}=-1+[(\om-\La_W) \pm \sqrt{\omega (\om-2 \La_W)}] \hat{r}^2 .
\label{Static_solution}
\end{\eq}
Then, one can check that the non-rotating metric (\ref{Static_metric}) transforms to the rotating metric (\ref{KAdS4}) under the coordinate transformation
\beq
\hat{t}=\f{t}{\Xi},~ \hat{r}^2=\f{1}{\Xi} (\De_{\theta} r^2-a^2 \sinh^2\theta),~\hat{r} \cosh \hat{\theta}=r \cosh \theta,~\hat{\phi}=\phi-\f{\La at}{3 \Xi},
\label{Diff_F}
\eeq
which can be considered as the
${\it Diff}_{\cal F}$ (\ref{deltx_Horava}-\ref{delg3})
\beq
\hat{t}=\hat{t} (t), ~\hat{\phi}=\hat{\phi} (\phi,t), ~\hat{r}=\hat{r} (r,\theta),~\hat{\theta}=\hat{\theta} (\theta, r).
\eeq

It is interesting to note that the above transformation is exactly the same as
(with the corresponding horizon topology)
in GR  with $\La$ in (\ref{Lambda_eq}) \ci{Cart:73,Henn}.
This may explain the solution of the cosmological constant $\La$ in (\ref{Lambda_eq}) as that of (\ref{Static_solution}) with the static metric
\beq
\hat{N}^2=\hat{f}=-1-\f{\La}{3} \hat{r}^2
\eeq
which has the same horizon as (\ref{horizon}), $\hat{r}_H=\sqrt{3/(-\La)}$,
and the Hawking temperature
\beq
\hat{T}_H=\f{\hat{f}'|_{r_H}}{4 \pi}=\f{\sqrt{-\La}}{2 \pi \sqrt{3}},
\label{Hawking_Temp}
\eeq
in agreement with (\ref{Hawking_Temp_H}). On the other hand, we note that the ergosphere in the rotating metric (\ref{ergo_KAdS}) is not preserved under the transformation, which may not be strange because of the non-inertial (rotating) property of the transformation (\ref{Diff_F}).
This can be compared with the appearance
of the {\it Rindler}
horizon in accelerating (Rindler) coordinates.

Similarly, one can check that the {zero-mass} $KdS_4$ metric can be also obtained by the corresponding ${\it Diff}_{\cal F}$ which can be obtained by the analytic continuation (\ref{analytic}) from (\ref{Diff_F}).
Moreover, due to the independence of {$\la$, as well as $\mu, \nu$, in} the solution
(\ref{Lambda_eq}),
one can discover that the different {massive} solutions with different $\la$ ($\la \neq 1/3$) \ci{Lu,Kiri} merge at the {zero-mass} solution. {The parameter independence of the {zero-mass} solution would be a natural result since it is just an indication of the {zero-mass} solution as the simplest configuration, {\it i.e.}, vacuum state, in the solution space. For example, the massive spherically symmetric solutions have the dependence of $\la$, though they do not have the dependence of $\mu, \nu$ \cite{Keha,Park:0905,Lu,Kiri}. Similarly, we expect that the massive rotating solutions, which will be the most general configurations, would depend on $\mu, \nu$, as well as $\la$, generally.}

Here, it is important
to note that there is a 
{{\it positive}} Hawking temperature {for the topological $(k=-1)$ AdS black holes} as in
(\ref{Hawking_Temp}) {or (\ref{Hawking_Temp_H})}, which
{being a strong indication of} the existence of Hawking radiation even though its mass $M$ vanishes. This would indicate the
existence of {\it negative mass} black holes by loosing its mass {$\De M=E_{(1)}<0$
with the absorbtion of the negative energy $E_{(1)}<0$ from either a pair of virtual particles
just outside the event horizon}
{in the Hawking radiation},
or
{a pair of split particles in
the ergo region (\ref{ergo_KAdS}), (\ref{ergo_KdS})}
in the Penrose process,
as well as the usual positive mass black holes by {gaining mass $\De M>0$ from}
the accreting (ordinary) matters satisfying the energy conditions. {In GR limit, the gravitational stability is proved for the {zero-mass} topological black hole and a stability condition is found for a negative-mass topological black hole as well \cite{Birm:2007}. It would be an open question whether the Lorentz-violating effects of \Ho~gravity changes the stability results for the zero or negative-mass topological black hole.}

It would be an outstanding open problem to find the {\it massive} rotating black holes
from the {zero-mass} rotating solution as a {\it seed} solution via {\it Kerr-Schild}-like method{, which treats the Kerr metric as the {zero-mass} rotating solution with the perturbed terms by its mass}
\ci{Kerr-Schild}.
{The main obstacle is that it is extremely difficult to find a proper ansatz
for the solvable differential equations, which are highly nonlinearly coupled
PDE in \Ho~gravity, generally. This is in contrast to the three-dimensional case where the
exact solutions for rotating black holes are obtained directly, without recourse to
Kerr-Schild-like method \cite{Park:1207}.}

{{\it Note added}: After finishing this paper, we recently studied a procedure for finding rotating black hole solutions in the low-energy sector of the four-dimensional (non-projectable) \Ho~gravity, and obtained the Kerr-type rotating black hole solutions with or without cosmological and electromagnetic charges, for the first time \cite{Deve:2024}. Without cosmological constant, similarly to Kerr-Schild method \cite{Kerr-Schild}, we introduced three mass-dependent metric ansatz functions into the {zero-mass} seed rotating solution and, after trails and errors, determined the undetermined functions by solving
{\it ODE}.
On the other hand, with a cosmological constant, it was a bit more complicated and we needed a more clever choice of the ansatz to find the non-trivial rotating solutions (for the details, see \cite{Doran-like}).  }

\section*{Acknowledgments}

We would like to thank D.~O.~Devecioglu for helpful correspondences.
This work was supported by Basic Science Research Program through the National
Research Foundation of Korea (NRF) funded by the Ministry of Education,
Science and Technology (NRF-2020R1A2C1010372, 2020R1A6A1A03047877 (MIP), NRF-
2018R1D1A1B05049338 (HWL)).
\newcommand{\J}[4]{#1 {\bf #2} #3 (#4)}
\newcommand{\andJ}[3]{{\bf #1} (#2) #3}
\newcommand{\AP}{Ann. Phys. (N.Y.)}
\newcommand{\MPL}{Mod. Phys. Lett.}
\newcommand{\NP}{Nucl. Phys.}
\newcommand{\PL}{Phys. Lett.}
\newcommand{\PR}{Phys. Rev. D}
\newcommand{\PRL}{Phys. Rev. Lett.}
\newcommand{\PTP}{Prog. Theor. Phys.}
\newcommand{\hep}[1]{ hep-th/{#1}}
\newcommand{\hepp}[1]{ hep-ph/{#1}}
\newcommand{\hepg}[1]{ gr-qc/{#1}}
\newcommand{\bi}{ \bibitem}

\end{document}